\documentclass[reqno,11pt]{amsart}

\IfFileExists{mymtpro2.sty}{%
  \usepackage[subscriptcorrection]{mymtpro2}
}{}

\usepackage{a4,amssymb,physics}

\usepackage{appendix}

\newtheorem{theorem}{Theorem}[section]
\newtheorem{lemma}{Lemma}[section]

\newtheorem{remark}[theorem]{Remark}

\newcommand{\labelnummer}{\mbox{\normalfont (\roman{numcount})}}%

\makeatletter

  {\let\curlabelspeicher\@currentlabel%
    \begin{list}{\labelnummer}%
      {\usecounter{numcount}\leftmargin0pt%
        \topsep0.5ex\partopsep2ex\parsep0pt\itemsep0ex\@plus1\p@%
        \labelwidth2.5em\itemindent3.5em\labelsep1em%
      }%
    \let\saveitem\item%
    \def\item{\saveitem%
      \def\@currentlabel{{\upshape\curlabelspeicher}$\,$\labelnummer}}%
    \let\savelabel\label%
    \def\label##1{\savelabel{##1}%
      \@bsphack%
        \ifmmode\else%
          \protected@write\@auxout{}%
          {\string\newlabel{##1item}{{\labelnummer}{\thepage}}}%
        \fi%
      \@esphack%
    }%
  }{\end{list}}%

\renewcommand{\appendix}{\def\thesection{\textsc{Appendix}}}

 \let\leq\le
 \let\geq\ge

\let\Re\undefined 
\DeclareMathOperator{\Re}{Re}

\makeatletter

\newif\ifper\pertrue
\def\per{.}

\def\bti{\@ifnextchar[\bbti\bbbti}
\def\bbti[#1]#2{#2, #1.}
\def\bbbti#1{#1.}

\def\z{\@ifnextchar[\zz\zzz}
\def\zz[#1]#2#3#4#5{\perfalse\emph{#2} \textbf{#3}, #4 (#5) [#1]}
\def\zzz#1#2#3#4{\emph{#1} \textbf{#2}, #3 (#4)\ifper\per\fi\pertrue}

\def\pub{\@ifstar\pubstar\pubnostar}
\def\pubnostar{\@ifnextchar[\@@pubnostar\@pubnostar}
\def\@@pubnostar[#1]#2#3#4{#2, #3, #4, #1\ifper\per\fi\pertrue}
\def\@pubnostar#1#2#3{#1, #2, #3\ifper\per\fi\pertrue}
\def\pubstar[#1]#2#3#4{\perfalse #2, #3, #4 [#1]\pertrue}

\makeatother

\newcommand{\cC}{{\mathcal C}}

\newcommand{\E}{{\mathcal E}}

\newcommand{\beq}{\begin{equation}}
\newcommand{\eeq}{\end{equation}}
\newcommand{\bea}{\begin{eqnarray}}
\newcommand{\eea}{\end{eqnarray}}

\newcommand{\R}{\mathbb{R}}

\newcommand{\N}{\mathbb{N}}

\newcommand{\C}{\mathbb{C}}

\newcommand{\bel}{\begin{equation} \label}
\newcommand{\ee}{\end{equation}}

{

\newcommand{\SCHR}{SCHR\"ODINGER }
\newcommand{\Schr}{Schr\"odinger }

\newcommand{\cF}{\mathcal{F}}

\newcommand{\rd}{\mathrm{d}}

\def\P{I\kern-.30em{P}}
\def\E{I\kern-.30em{E}}
\renewcommand{\E}{\mathbb{E}\mkern2mu}
\renewcommand{\P}{\mathbb{P}}

\begin{document}

\title[Edge currents for time-fractional \SCHR equations]{Edge currents for the time-fractional, half-plane, \SCHR equation with constant magnetic field}

\author[P.\ D.\ Hislop]{Peter D.\ Hislop}
\address{Department of Mathematics,
    University of Kentucky,
    Lexington, Kentucky  40506-0027, USA}
\email{peter.hislop@uky.edu}

\author[\'E.\ Soccorsi]{\'Eric Soccorsi}
\address{Aix Marseille Univ, Universit\'e de Toulon, CNRS, CPT, 
Marseille, France}
\email{eric.soccorsi@univ-amu.fr}


\medskip

\begin{abstract}
We study the large-time asymptotics of the edge current for a family of time-fractional \Schr equations with a constant, transverse magnetic field on a half-plane $(x,y) \in \R_x^+ \times \R_y$. The TFSE is parameterized by two constants $(\alpha, \beta)$ in $(0,1]$, where $\alpha$ is the fractional order of the time derivative, and $\beta$ is the power of $i$ in the \Schr equation. We prove that for fixed $\alpha$, there is a transition in the transport properties as $\beta$ varies in $(0,1]$: For $0 < \beta < \alpha$, the edge current grows exponentially in time, for $\alpha = \beta$, the edge current is asymptotically constant, and for $\beta > \alpha$, the edge current decays in time. We prove that the mean square displacement in the $y\in \R$-direction undergoes a similar transport transition. These results provide quantitative support for the comments of Laskin \cite{laskin2000_1} that the latter two cases, $\alpha = \beta$ and $\alpha < \beta$, are the physically relevant  ones. 

\bigskip
\begin{center}
\emph{\large{Dedicated to the memory of our colleague and friend  Georgi Raikov}
}
\end{center}

\end{abstract}

\maketitle \thispagestyle{empty}



\tableofcontents



\section{Introduction and statement of the problem}\label{sec:introduction}
\setcounter{equation}{0}

There has been a lot of analysis of systems described by equations with space or time fractional derivatives, especially associated with parabolic and hyperbolic partial differential equations. In this article, we study the time-fractional \Schr equation (TFSE) for a physically motivated model. The TFSE studied in this paper has the form
\beq\label{eq:tfse_defn1}
i^\beta \partial_t^\alpha u (t,x) = H u(x,t),
\eeq
on the Hilbert space $L^2(\Omega)$, for certain self-adjoint \Schr operators $H$. The fractional time derivative is the Caputo derivative defined in \eqref{eq:cap1}. The parameters $\alpha, \beta \in (0,1]$. We show how the time-fractional derivative affects the large time behavior of the transport of the system. In particular, the model is described by a magnetic \Schr operator in a half-plane $\Omega := \R^+_x \times \R_y$, with a Dirichlet boundary condition at $x=0$. Such operators and geometry play a role in the integer quantum Hall effect. 

These models allow us to examine the affect of a fractional time derivative on the transport properties of the model. The dynamics of the \Schr equation ($\alpha = \beta =1$ in \eqref{eq:tfse_defn1}) exhibits nontrivial edge currents, whose conductance is quantized, and which flow along the 
impenetrable edge $x=0$ in the sense that they are localized in a small neighborhood of the edge. We now replace the dynamical \Schr equation by the time-fractional \Schr equation \eqref{eq:tfse_defn1} and examine the edge current associated with these half-space operators. Our main results are that the edge current exhibits two types of extreme behavior depending on the relation between the two parameters $(\alpha, \beta)$: either the current decreases to zero as $t \rightarrow \infty$, or the edge current explodes as $t \rightarrow \infty$. These two distinctly different behaviors depend on how the time-fractional \Schr equation is defined. At the critical values, the edge current is asymptotically constant, reminiscent of the \Schr case ($\alpha = \beta =1$) for which the edge current is constant.  


%
%

In this paper, we study a two-parameter family of TFSE and compute the asymptotics of the edge current and the mean square displacement. We show, for example, that the edge current of the Naber model \eqref{eq:naber0} is asymptotically a nonzero constant, whereas in the AYH model \eqref{eq:achar0}, it decays to zero. Furthermore, we study the mean-square displacement of the solutions in the 
$y$-direction and prove that for  the Naber model, this is asymptotically ballistic, whereas for the AYH model, it decays to zero.  
 
 
\subsection{Fractional quantum mechanics}\label{subsec:tfqm1}
 
Fractional quantum mechanics (FQM) was introduced by Laskin in 1999, and developed by him in a series of papers \cite{laskin2000_1,laskin2000_2,laskin2002}. The crucial component of FQM is the fractional \Schr equation (FSE). In the FSE, the normal Laplacian in the space variables, $-\Delta$, is replaced with a fractional Laplacian, defined through the functional calculus, $(-\Delta)^\alpha$, with $\alpha \in (0, 1]$:
$$ 
- i \partial_t u + (-\Delta)^\alpha  u + q u =0,\ \alpha \in (0,1] . 
$$
The fractional Laplacian is a spatially non-local operator. However, due to the presence of the first-order time derivative in the FSE, FQM preserves all quantum mechanics fundamentals such as the probability conservation law and the energy conservation law, as follows from the unitarity of the time-evolution operator.

The regular \Schr equation has the mathematical appearance of a diffusion equation and can be derived by considering probability distributions. Feynman and Hibbs \cite{FH} used a Gaussian probability distribution in the space of all possible paths for a quantum mechanical particle, in order to derive the \Schr equation. Laskin derived the FSE by using Feynman's path integral approach with L\'evy distributions, instead of Gaussian distributions, for the set of possible paths.

Naber \cite{naber} introduced the TFSE in 2004, 
by replacing the first-order time derivative in the \Schr equation with a fractional power $\alpha \in (0,1]$, and raising $i$ to the same power $\alpha$: 
\beq\label{eq:naber0}
- i^\alpha \partial_t^\alpha u - \Delta u + q u =0. 
\eeq
We will refer to this TFSE as the \textbf{Naber model.} The fractional time derivative is defined to be the Caputo derivative, see \eqref{eq:cap1}. 

One of the main reasons invoked by Naber for raising $i$ to the power of the time derivative is the asymptotic stability of the time-evolution of the solutions to the TFSE with respect to $\alpha$. By asymptotic stability, we mean that the infinite $t$ behavior  
is independent of the order $\alpha$ of the derivative, see, for example, \eqref{eq:asympt00}.

The Naber model was not derived from the classical \Schr equation by considering a non-Markovian evolution but the time-evolution of the solution to Naber's TFSE is reminiscent of the time-evolution of solutions to the regular \Schr equation. Nevertheless, the time-evolution of solutions in the Naber model for $0 < \alpha < 1$ is not unitary, so that it does not support the the basic quantum mechanical laws of probability conservation and energy conservation.

In 2013, Narahari Achar, Yale, and Hanneken \cite{ayh} provided another model of a TFSE that differs from the Naber model.  These authors presented a derivation that relies on a suitable representation of the action in the Feynman path integral motivated by the fractional dynamics of a free classical particle. In what we will call the \textbf{AYH model} of the TFSE, the operator $i \partial_t$ is replaced by $i \partial_t^\alpha$, with $\alpha \in (0,1)$:
\beq\label{eq:achar0} 
 - i \partial_t^\alpha u -\Delta u + q u =0. 
 \eeq
Like the Naber model, the time-evolution is not unitary, and, consequently, the AYH model has the same shortcomings in that it violates the basic quantum mechanical laws.

However, Narahari Achar, Yale, and Hanneken \cite{ayh} argued that this model is more physically natural than the one proposed by Naber. 
For example, the total probability associated with a solution  $\psi(x,t)$ of the \Schr equation or TFSE is given by $\int_{\R^d} | \psi(x,t) |^2 dx$. The unitary time evolution of the solutions to the usual \Schr equation guarantees that this quantity is equal to one for all time (with a normalized initial condition). But, for the models of both Naber and AYH, this quantity depends on time. As $t \rightarrow \infty$, the total probability is greater than one in the Naber model, whereas for the AYH model, the total probability vanishes. 

There have been several other papers on the TFSE since Naber's article \cite{naber} and we briefly mention the following works related to our results. Dong and Xu \cite{DX} (see also \cite{WX}) discussed a general space-time fractional \Schr equation combining the Caputo fractional time derivative with $\beta \in (0,  2)$, with a fractional space derivative of order $\alpha \in (0,2)$ (the notation of \cite{DX}). The resulting fractional space-time \Schr equation has the form
$$
i^\beta \partial_t^\beta u (x,t) = ((- \Delta)^\frac{\alpha}{2} u)(x,t) + V(x) u(x,t)
= (H_\beta u)(x,t),
$$
on the Hilbert space $L^2(\R^d)$  with an initial condition $u(0) = u_0$. 
The fractional Laplacian $(- \Delta)^\frac{\alpha}{2}$ is defined through the Fourier transform as the multiplication operator by $|k|^{\alpha}$. 

Dong and Xu \cite{DX} studied the impact of the potential $V$ on the time evolution of the solutions with initial conditions chosen to be eigenfunctions of the Hamiltonian $H_\alpha := (- \Delta)^\frac{\alpha}{2} + V$. With this choice, they find product solutions for which the time dependence is expressed through a Mittag-Leffler function. They determined that the long time behavior depends on the order $\beta$ of fractional time derivative and the sign of the eigenvalue of $H_\alpha$ corresponding the the chosen eigenfunction for the initial condition. In particular, if the eigenvalue is positive, and  $0 < \beta \leq 1$, the total probability of the solution never tends to zero, whereas if $1 < \alpha < 2$, the total probability of a solution tends to zero in certain cases.

Bayin \cite{bayin} also explored the relation between the eigenvalues of $H_\alpha$ and the large-time behavior of the solutions. He studies solutions of the form $u(x,t) = X(x)T(t)$. Assuming the spectrum of $H_\alpha$ is discrete, as in \cite{DX}, the function $X$ is an eigenfunction of $H_\alpha$. The temporal part $T$ is given by a Mittag-Leffler function determined by $\beta$. Bayin proved the nonconservation of probability. He also gave the long-time behavior of product solutions to the TFSE following from the asymptotics of the Mittag-Leffler functions. There are similarities between these papers and our treatment in section \ref{sec:well_posed1}. 

P.\ G\'orka, H.\ Prado,  and J.\ Trujillo \cite{GPT1} studied the abstract Naber model $i^\alpha \partial_t^\alpha u = H u$, $\alpha \in (0,1)$ for a general nonnegative, self-adjoint operator $H$ on a Hilbert space $\mathcal{H}$. They prove existence and uniqueness of strong solutions in $\mathcal{H}$ when $\partial_t^\alpha$ is the Caputo derivative, and the initial condition $u_0 \in D(H) \subset \mathcal{H}$. Among their results, they prove that the solution operator $U_\alpha(t)$ approaches the unitary time evolution $e^{-itH}$ as $\alpha \rightarrow 1^-$. Their approach, based properties of the Mittag-Leffler functions on the spectral theorem for $H$, has some similarities with ours, see section \ref{sec:well_posed1}. 

P.\ G\'orka, H.\ Prado,  and D.\ J.\ Pons \cite{GPP1} continued this study and examined the time-asymtotic behavior of solutions to the abstract Naber model. 
They also proved that the global probability is not preserved under the evolution operator $U_\alpha(t)$: If $u_0 \in \mathcal{H},$ with $\| u_0 \|_{\mathcal{H}} = 1$, then $\| U_\alpha(t) u_0\|_{\mathcal{H}}$ is $\mathcal{O}(\frac{1}{\alpha})$. 
This had already been noted by \cite{DX}.

A.\ Iomin \cite{iomin2009,iomin2011} also studied the Naber model \eqref{eq:naber0} in relation to the quantum comb model. He used an eigenfunction expansion of the spatial Hamiltonian and the Mittag-Leffler functions to express the Green's function as 
$$
G(x,x',t) = \sum_{\lambda} \phi_\lambda^*(x') \phi_\lambda (x) E_{\alpha} ( \lambda (-i t)^\alpha).
$$
For $\alpha = \frac{1}{2}$, Iomin discussed similarities between the Green's function for the TFSE and the quantum comb model. 
In \cite{iomin2019_A}, Iomin studied the AYH model \eqref{achar0} and explored quantum mechanical properties of the model, such as the Heisenberg equations of motion. 
A review of TFSE is given in \cite{iomin2019_B}.

\subsection{The quantum Hall model}\label{subsec:model1}

We study the magnetic \Schr operator in a half-plane with a constant, transverse magnetic field of strength $b > 0$ and Dirichlet boundary conditions along the edge $x=0$. The half-plane is $\Omega := \{ (x,y) ~|~ x > 0, y \in \R \}$. The constant magnetic field is orthogonal to the half-plane and described by a vector potential  $\textbf{a}(x,y) := (a_1(x,y), a_2(x,y)) := b (0,x)$.  

The magnetic \Schr operator is defined as follows. Let $p_x := -i \partial_x$ and $p_y := - i \partial_y$ be the two momentum operators. The two-dimensional magnetic Schr\"odinger operator $H(\textbf{a})$
is defined on the dense domain ${\rm C}_0^\infty (\Omega) \subset {\rm L}^2 (\Omega)$ by
\bel{eq:sharp1}
H=H(\textbf{a}) := (-{\rm i} \nabla - \textbf{a})^2 = p_x^2 + (p_y + bx)^2.
\ee

The spectral properties of $H(\textbf{a})$ are well-known. 
These are derived from the representation of $H(\textbf{a})$ as a direct integral. 
Because of the translational invariance in the $y$-direction, the partial Fourier transform with respect to $y$ provides this decomposition. 
Let $\cF_y$ be the partial Fourier transform with respect to $y$, i.e.
$$
\hat{\varphi}(x,k) = (\cF_y \varphi)(x,k) :=  \frac{1}{\sqrt{2 \pi}} \int_{\R} e^{-ik y} \varphi(x,y) \rd y,\ \varphi \in L^2(\Omega).
$$
The direct integral decomposition of $H(\textbf{a})$ is given by
\bel{fiber-dec}
\cF_y H(\textbf{a}) \cF_y^* = \int_\R^\oplus h_b(k) ~dk,
\ee
where the reduced \Schr operator $h_b(k)$ is given by
\bel{D:ghk}
h_b(k):=-\partial_{x}^2+ (k+bx)^2,\ x>0,\ k \in \R .
\end{equation}
These operators act on $L^2(\R^{+})$ with Dirichlet boundary condition at $x=0$. 


For all $k \in \R$ fixed, the effective potential $V_k(x) := (k+bx)^2$ is unbounded as $x$ goes to infinity, so $h_b(k)$ has a compact resolvent. Let $\{ \lambda_n(k),\ n \in \N \}$ denote the eigenvalues, arranged in non-decreasing order, of $h_b(k)$. Since all these eigenvalues $\lambda_n(k)$ with $n \in \N$ are simple, each $k \mapsto \lambda_n(k)$ is a real analytic function in $\R$.
The \emph{dispersion curves} $\lambda_{n}(k)$, $n \in \N$, have been extensively studied in several articles (see e.g. \cite{DeB-P,hs1}). They are monotone decreasing functions of $k \in \R$, obeying
\bel{E:limband}
\lim_{k\to-\infty}\lambda_{n}(k)=+\infty \ \  \mbox{and}\  \ \lim_{k\to+\infty}\lambda_{n}(k)=E_{n},
\end{equation}
for all $n \in \N$, where $E_n:= (2n-1)b$ is the $n$-th Landau level.
As a consequence, the spectrum of $H(\bf{a})$ is $\sigma(H({\bf{a}}))=\overline{\cup_{n\geq1}\lambda_{n}(\R)}=[b,+\infty)$.
The Landau levels $E_{n}$, $n \in \N$, are thresholds in the spectrum of $H(\bf{a})$.

%

For $n \in \N$ and $k \in \R$, we consider a normalized eigenfunction $\phi_{n}(\cdot,k)$ of $h_b(k)$ associated with $\lambda_{n}(k)$. It is well known that $\phi_{n}(\cdot,k)$ depends analytically on $k$. We define the $n$-th generalized Fourier coefficient of $u \in L^2(\Omega)$, by
\bel{def-phin}
u_{n}(k):=\langle \cF_{y} u(\cdot,k),\phi_{n}(\cdot,k) \rangle_{L^2(\R^{+})}= \frac{1}{\sqrt{2\pi}}\int_{\R^{+}}\hat{u}(x,k) \overline{\phi_{n}(x,k)} ~dx. 
\ee
Note that the inner product in \eqref{def-phin} is linear in the first entry. 
In this setting, Parseval's Theorem yields
\bel{eq:l2-prop1}
\| u \|_{L^2 (\Omega)}^2 = \| \widehat{u} \|_{L^2(\Omega)}^2 = 
\sum_{n\geq1}  \| u_n \|_{L^2(\R)}^2.
\ee



\subsection{The edge current}\label{subsec:current1}

We define the \emph{edge current} carried by a state as the time derivative of the 
expectation of  the observable $y$ in the time-evolved state. Here, the observable $y$ denotes the multiplier by the coordinate $y$ in $L^2(\Omega)$, with $\Omega = \R^+_x \times \R_y$. Given an evolution equation, we denote by $u(t)$ the dynamical solution to the initial value problem with $u(t=0) = u_0$. The edge current $J_y[u_0](t)$ is given by
\beq\label{eq:current1}
J_y[u_0](t) := \frac{d}{dt} \langle u (t), y u(t) \rangle_{L^2(\Omega)} . 
\eeq

For the ordinary \Schr equation, we have $u(t) = e^{-itH}u_0$.
In this case, the time evolution can be represented by the Heisenberg variable
$y(t):= e^{-itH} y e^{itH}$,  all $t \in \R$. A current operator may be expressed in terms of the time derivative of $y(t)$. This is the velocity in the $y$-direction. As is well known,  this operator is given by
$ \frac{\rd y(t)}{\rd t} = -i [ H, y(t)]= -i e^{-i tH} [H, y] e^{itH}$.
In this case, the current operator has the form of the self-adjoint operator
$ J_{y}:=-i [H,y]= -2 (p_y + bx)$. 
The current carried by a time-evolved state $u(t)$, with initial condition $u_0$, can the be written in terms of the current operator: 
\bea\label{eq:current2}
J_y[u_0](t) & := & \frac{d}{dt} \langle u(t), y u(t) \rangle_{L^2(\Omega)} =
 -i \langle u(t), [H,y] u(t) \rangle_{L^2(\Omega)}   \nonumber \\
  & = & \langle u(t), J_y u(t) \rangle_{L^2(\Omega)}  = \langle u_0 , J_y(t) u_0 \rangle_{L^2(\Omega)}   ,  
\eea
where the Heisenberg current observable is defined by 
$J_y(t) := e^{-itH} J_y e^{itH}$.

For the TFSE \eqref{eq:tfse_defn1}, the time evolution is not given by a unitary group. As discussed below, the time evolution is expressed, via the functional calculus, in terms of a Mittag-Leffler function. These functions, whose properties are summarized in Appendix \ref{app1:ml1}, are defined by 
\beq\label{eq:defn_ml1}
E_{\alpha, \sigma}(z) := \sum_{k=0}^\infty \frac{z^k}{\Gamma ( \sigma + \alpha k)} ,
\eeq
where $\Gamma (z)$ denotes the gamma function. It follows from \eqref{eq:defn_ml1}  that $E_{1,1}(z) = e^z$. 
In terms of these functions, the solution to the TFSE may be written as
\beq\label{eq:tfse_soln0}
u(t) = E_{\alpha,1}((-i)^\beta t^\alpha H)u_0.
\eeq
The current then has the form
\beq\label{eq:current3}
J_y[u_0](t) := \frac{d}{dt} \langle u (t), y u(t) \rangle_{L^2(\Omega)} = \frac{d}{dt} \int_{\Omega} y | (E_{\alpha,1}((-i)^\beta t^\alpha H)u_0) (x,y)|^2 ~dx ~dy . 
\eeq


\subsection{Summary of results on the edge current}\label{subsec:results1}

The dynamics of the TFSE is governed by the initial-value problem:
\beq\label{eq:dyn1}
i^\beta \partial_t^\alpha u(x,y;t) = H(\textbf{a}) u (x,y;t) =(p_x^2 +(p_y + bx)^2)u(x,y,t),
\eeq
with the initial condition $u(t=0) = u_0(x,y).$  The solution to the initial-value problem is
$$
u(t) = E_{\alpha,1}((-i)^\beta  t^\alpha H(\textbf{a}))u_0 ,
$$
see \eqref{eq:tfse_soln0} and \eqref{eq:defn_ml1}. 
The exponents $(\alpha, \beta) \in (0, 1] \times [0,1]$. There are different dynamical asymptotic regimes depending on the relationship between these two exponents. Our general results are:

\begin{enumerate}
\item \textbf{Case 1:} $0 < \beta \leq \alpha < 1$. 
For $\beta < \alpha$, the edge current grows exponentially in time. In the case $0 < \alpha = \beta < 1$, the edge current is asymptotically constant in time.

\medskip

\item \textbf{Case 2:} $0 < \alpha < \beta \leq 1$. The current exhibits
decays to zero as $t \rightarrow \infty$ like $t^{-1- 3 \alpha}$. 

\end{enumerate}

%
%

\medskip

In particular, we mention three special cases discussed in the literature:

\medskip

\begin{enumerate}
\item\textbf{ \Schr equation} \cite{hs1}: $\alpha=\beta = 1$. The solution to the initial-value problem is
$$
u(t) = E_{1,1}(-i t H(\textbf{a}))u_0 = e^{-it H(\textbf{a})} u_0.
$$
The edge current is bounded for all time, see Remark \ref{rmk:schr_case1}. Lower bounds on the edge current were derived in \cite{hs1}. 

\medskip

\item\textbf{ Naber model} \cite{naber}: $\alpha = \beta \in (0,1)$. The solution to the initial-value problem is
$$
u(t) = E_{\alpha,1}((-i)^\alpha t^\alpha H(\textbf{a}))u_0.
$$
The edge current is bounded for all time. This is similar to the \Schr case. 

\medskip

\item \textbf{AYH model} \cite{ayh}: $\alpha \in (0,1)$ and $\beta = 1$. The solution to the initial-value problem is
$$
u(t) = E_{\alpha,1}(-i t^\alpha H(\textbf{a}))u_0.
$$
The edge current decays to zero like $t^{-1- 3 \alpha}$.

%
\end{enumerate}

\medskip

%
%
%
%
%

These results illustrate that, for fixed $0 < \alpha < 1$, tuning the parameter $\beta \in (0,1]$ leads to qualitatively different behaviors in the long-time asymptotics of the current. The critical value $\beta = \alpha$ is a transition between exponential growth and inverse power decay. 
For $0 < \alpha < 1$ \textbf{fixed}, as $\beta \in (0,1)$ varies, we have 

\begin{eqnarray}\label{eq:asympt00} 
0< \beta < \alpha & J_y[u_0](t) \sim e^{ct} \nonumber \\ 
  &    \nonumber \\ 
  \beta = \alpha &  J_y[u_0](t) \sim C_0  \nonumber   \\ 
      &   \nonumber  \\  
  \alpha < \beta \leq 1  &  J_y[u_0](t) \sim t^{-1-3 \alpha}  \nonumber \\ 
  \end{eqnarray}
  
  \medskip
  
 The \Schr case $\alpha = \beta = 1$ is conservative and the edge current is constant in time. For other values of  $\alpha = \beta$, the current is asymptotically constant. 
 

\subsection{Outline of the paper}

In section \ref{sec:well_posed1}, we prove the existence and uniqueness of the solution to the initial-value problem with $u_0$ chosen from an appropriate function space depending on the exponents $(\alpha, \beta)$. Section \ref{sec:current1} presents the calculation of the edge current $J_y[u_0](t)$ in terms of the Mittag-Leffler functions. The main results on the asymptotic behavior of the edge currents, for various values of the exponents  $(\alpha, \beta)$, are presented in section \ref{sec:asympt1}, including a the special cases of the models of Naber and of AYH. In section \ref{sec:msd1}, we calculate the mean-square displacement (MSD) of the $y$-observable for the time-fractional Hall model for the ranges of exponents  $(\alpha, \beta)$. Details concerning the Mittag-Leffler functions are summarized in the Appendix \ref{app1:ml1}.


\subsection{Acknowledgement} The authors thank Yavar Kian for discussions on the topics of this paper.  PDH thanks Aix Marseille Universit\'e for some financial support and hospitality during the time parts of this paper were written. PDH is partially supported by Simons Foundation Collaboration Grant for Mathematicians No.\ 843327. ÉS is partially supported by the Agence Nationale de la Recherche (ANR) under grant ANR-17-CE40-0029.

\section{The well-posedness of constant-order time-fractional Schr\"odinger equations}
\label{sec:well_posed1}
\setcounter{equation}{0}

We prove the existence and uniqueness of the solution to the initial-value problem for the time-fractional \Schr equation (TFSE) on general general subdomains $\Omega_0 \subset \R^d$, $d \geq 1$, with smooth boundary $\Gamma_0:=\partial \Omega_0$. For $\alpha, \beta  \in (0,1)$,
the initial-value problem for the time-fractional \Schr equation is:
\beq\label{eq:ivp_fs1}
- i^\beta \partial_t^\alpha u(\cdot,t) + H u(\cdot,t) = 0 , ~~~\mbox{and} ~~~u(\cdot,t=0) = u_0(\cdot).
\eeq

We assume that $\Omega_0$ is open and not necessarily bounded. Let $H$ be a lower semi-bounded (nonnegative) self-adjoint operator with domain $D(H) \subset L^2(\Omega_0)$ and having discrete spectrum. For the sake of simplicity, we assume that $D(H^{\frac{1}{2}})=H_0^1(\Omega_0)$, and that $H$ has a compact resolvent. 
We denote by $\lambda_1 \leq \lambda_2 \leq \ldots$ the eigenvalues of $H$ arranged in nondecreasing order and repeated with the multiplicity. We pick an orthonormal basis $\{ \phi_n,\ n \in \N \}$ in $L^2(\Omega_0)$ of eigenvectors of $H$, satisfying
$$ H \phi_n = \lambda_n \phi_n. $$

In order to use the results of this section, we recall that the quantum Hall model is described on $\Omega = \R^+_x \times \R_y$, with the self-adjoint \Schr operator $H({\bf{a}})$ on $L^2(\R^+_x \times \R_y)$. By means of the partial Fourier transform with respect to $y$, see \eqref{fiber-dec}, the problem is reduced to the study of the family of fiber operators $h_b(k)$, $k \in \R$, acting on $L^2(\R_x^+)$. These nonnegative operators $h_b(k)$ have compact resolvent and, consequently, discrete, simple spectrum.  
So, the results of this section apply to the operators $h_b(k)$ with $\Omega_0 = \R^+_x$.

\subsection{Main results on the TFSE}\label{subsec:tfse_results1}

We recall the time-fractional Schr\"odinger equation for exponents $(\alpha, \beta) \in (0,1) \times (0,1)$. We will also consider the endpoints $\alpha, \beta \in \{0, 1\}$ separately. 
The TFSE is the initial-value problem 
\bel{eq1} 
-i^\beta \partial_t^\alpha u(x,t) + H u(x,t) = 0,\ (x,t) \in \Omega_0 \times (0,T)
\ee
with homogeneous Dirichlet boundary condition
\bel{eq2}
u(x,t) = 0,\ (x,t) \in \Gamma_0  \times (0,T)
\ee
and initial state $u_0$
\bel{eq3}
u(x,0) = u_0 (x),\ x \in \Omega_0,
\ee
where $\partial_t^\alpha$ denotes the Caputo fractional derivative of order $\alpha$. The Caputo fractional derivative $\partial_t^\alpha$ is defined by 
\beq\label{eq:cap1}
\partial_t^\alpha u(t)=\frac{1}{\Gamma(1-\alpha)} \int_0^t \frac{u'(s)}{(t-s)^{\alpha}} ds.
\eeq
We note that $\partial_t^\alpha u \rightarrow \partial_t u$, as $\alpha \rightarrow 1$,  if, for example, $u \in C^2(\R)$.  This follows from the formula \eqref{eq:cap1} by an integration by parts. 

We call $u(x,t)$, with $x \in \Omega_0$ and $t > 0$, a \textbf{solution} to \eqref{eq1}--\eqref{eq3} if the 3 following conditions hold simultaneously:
\begin{enumerate}
\item $u(\cdot,t) \in D(H)$, for a.e. $t \in \R^+$;
\item the fractional \Schr equation \eqref{eq1} holds in $L^2(\Omega_0)$ for $t \in \R^+$:
$$
i^\beta \partial_t^\alpha u(\cdot,t) =  H u(\cdot,t) ;
$$ 
\item $u \in \cC(\overline{\R^+},L^2(\Omega_0))$, so $\lim_{t \downarrow 0} \norm{u(\cdot,t)-u_0}_{L^2(\Omega_0)}=0$.
 \end{enumerate}
 
 Using an eigenfunction expansion, we prove the existence and uniqueness of solutions to the TFSE. We need to take the initial conditions from various function spaces in $L^2 (\Omega_0)$ depending on the relationship between $\alpha$ and $\beta$. Recall that $D(H)$ denotes the domain of the self-adjoint operator $H$ in $L^2 (\Omega_0)$. For an orthonormal basis $\{ \phi_n \}_{n \in \N }$ and $u_0 \in L^2 (\Omega_0)$, we define the associated Fourier coefficients by $u_{0,n} := \langle u_0, \phi_n \rangle_{L^2 (\Omega_0)}$. 
 We denote by $C_{F}(\Omega_0)$ the dense set of initial conditions $u_0 \in L^2(\Omega_0)$ with finitely-many nonzero Fourier coefficients $u_{0,n}$.

\begin{theorem}
\label{thm:existence1}
Let $(\alpha, \beta) \in (0,1)^2$ be the parameters of the TFSE described in \eqref{eq1}-\eqref{eq3}. We distinguish three cases: 
\begin{enumerate}
\item Case 1: $0 < \alpha < \beta \leq 1$. Let $u_0 \in L^2(\Omega_0)$. 

\medskip

\item Case 2: $0< \alpha = \beta \leq 1$. Let $u_0 \in D(H)$.

\medskip

\item Case 3: $0 \leq \beta < \alpha \leq 1$. Let $u_0 \in C_F(\Omega_0)$. 

\end{enumerate}

\medskip
\noindent
In each case, there exists a unique solution $u \in C(\overline{\R^+},L^2(\Omega_0)) \cap C(\R^+,D(H))$ to \eqref{eq1}--\eqref{eq3} of the form
\bel{eq4}
u(x,t)=\sum_{n=1}^{\infty} u_{0,n} E_{\alpha,1}((-i)^\beta \lambda_n t^\alpha) \phi_n(x),\ u_{0,n}:=\langle u_0 , \phi_n \rangle_{L^2(\Omega_0)},
\ee
such that 
\begin{enumerate}
\item Case 1: For each $u_0 \in L^2 (\Omega_0)$, there exists a finite constant 
$C_1 \geq 0$,
so that  
$$
\norm{u}_{\cC({\R^+},L^2(\Omega_0))} \leq C_1 \norm{u_0}_{L^2(\Omega_0)}.
$$
Moreover, we have $\partial_t^\alpha u \in C(\R^+,L^2(\Omega_0))$, and there exists a finite constant $\tilde{C}_1 > 0$ so that 
$$
 \norm{u(\cdot,t)}_{D(H)} + \norm{\partial_t^\alpha u(\cdot,t)}_{L^2(\Omega_0)} \leq \tilde{C}_1 \norm{u_0}_{L^2(\Omega_0)} t^{-\alpha}, 
$$
for all $t > 1$. 

\medskip
\item Case 2: For each $u_0 \in D(H)$, there exists a finite constant $C_2 \geq 0$,
so that 
$$
\norm{u}_{\cC(\overline{\R^+},L^2(\Omega_0))} \leq C_2 \norm{u_0}_{D(H)} .
$$
Moreover, we have $\partial_t^\alpha u \in C(\R^+,L^2(\Omega_0))$, and
there exists a finite constant $\tilde{C}_2 \geq 0$,
so that 
$$
\norm{u(\cdot,t)}_{D(H)} + \norm{\partial_t^\alpha u(\cdot,t)}_{L^2(\Omega_0)} \leq 
\tilde{C}_2 \norm{u_0}_{D(H)}, \forall t \in \R^+.
 $$
\medskip

\item Case 3: Let $\theta (\alpha, \beta) := \frac{\pi \beta}{2 \alpha}$.  For each $u_0 \in C_F(\Omega_0)$, there exists a finite constant $C_3 \geq 0$, and an index $n^* \in \N$, so that 
$$
\norm{u}_{\cC(\overline{\R^+},L^2(\Omega_0))} \leq C_3 (1+ t^\alpha \lambda_{n^*})^{2 \left( \frac{1-\beta}{\alpha} \right)} e^{2t \lambda_{n^*}^{\frac{1}{\alpha}}\cos \theta(\alpha,\beta)}  \| u_0 \|_{L^2(\Omega_0)}^2
$$
Moreover, we have $\partial_t^\alpha u \in C(\R^+,L^2(\Omega_0))$ and there exists a finite constant $\tilde{C}_3 \geq 0$,
so that
\bea
 \lefteqn{\norm{u(\cdot,t)}_{D(H)} + \norm{\partial_t^\alpha u(\cdot,t)}_{L^2(\Omega_0)}  } \nonumber \\
  & \leq & \tilde{C}_3 (1+ t^\alpha \lambda_{n^*})^{2 \left( \frac{1-\beta}{\alpha} \right)} e^{2t \lambda_{n^*}^{\frac{1}{\alpha}}\cos \theta(\alpha,\beta)}
    \| u_0 \|_{L^2(\Omega_0)}^2 .
 \eea
 \end{enumerate}
 The constants $C_j$, $j=1,2,3$ depend on $\alpha, \beta, \Omega$, and $H$. 
\end{theorem}

%
%
%
%


\subsection{Proof of Theorem \ref{thm:existence1}}\label{subsec:proof1}

We divide the proof of Theorem \ref{thm:existence1} into two parts:  existence and uniqueness. 

\begin{proof}
\noindent \emph{Part 1: \textit{Existence}}.

\noindent 
1. We show that the solution exists as the $L^2$-norm limit of the sequence of partial sums given by the expansion of the solution with respect to the orthonormal basis $\phi_n$ of eigenfunctions of $H$. The relationship between the exponents $\alpha$ and $\beta$ determine the choice of the initial conditions $u_0$ for which these expansion converge. This is controlled by the bounds on the Mittag-Leffler functions as described in section \ref{subsec:MLbounds1}. 
Given $u_0 \in L^2 (\Omega_0)$, we let $u_{0,n}:=\langle {u_0}, \phi_n \rangle_{L^2(\Omega_0)}$ be the $n^{th}$-Fourier coefficient of $u_0$ relative to this orthonormal basis. We first consider the finite sum
\beq\label{eq:sum1}
S_N(x,t) := \sum_{n=1}^{N} u_{0,n} E_{\alpha,1}((-i)^\beta \lambda_n t^\alpha) \phi_n(x), ~~~N \in \N  . 
\ee
We define the function $F_{\alpha,\beta,n}(t)$, for $t \geq0$, based on the bounds \eqref{eq:asymp1} and \eqref{eq:asymp2}:
\begin{eqnarray}\label{eq:bound1}
0 < \alpha < \beta &  F_{\alpha,\beta, n}(t) = \frac{C}{1+t^\alpha \lambda_n}  \leq C_1  \label{eq:case1} \\
 & & \nonumber  \\
\alpha = \beta & F_{\alpha,\beta, n}(t) = C_2 \label{eq:case2} \\
 & & \nonumber \\
 \beta < \alpha \leq 1 & F_{\alpha,\beta, n}(t) = C_3 \left[ (1+ t^\alpha \lambda_n)^{\frac{1-\beta}{\alpha}} e^{t \lambda_n^{\frac{1}{\alpha}}\cos \theta(\alpha,\beta)} + \frac{1}{1 + t^\alpha \lambda_n} \right] , \label{eq:case3} \nonumber \\
  & & \nonumber \\
 \end{eqnarray}
where  $ \theta(\alpha, \beta) := \left( \frac{\pi \beta}{2 \alpha} \right) $.  The finite constants $C_j> 0$, for $j=1,2,3$,  depend on $(\alpha, \beta)$, but are independent of $(n,t)$. 

\noindent
2. For any $t \geq 0$, we have
\begin{eqnarray}\label{eq:s_n1}
\norm{S_N(t)}_{L^2(\Omega_0)}^2 & = & \sum_{n=1}^{N} \abs{u_{0,n}}^2 \abs{E_{\alpha,1}((-i)^\beta \lambda_n t^\alpha)}^2 \nonumber \\
& \leq &  \sum_{n=1}^{N} \abs{u_{0,n}}^2 F_{\alpha,\beta,n}^2(t).
\end{eqnarray}
For cases 1 and 2, the bounds \eqref{eq:case1} and \eqref{eq:case2} show that the function $F_{\alpha,\beta,n}(t)$ is uniformly bounded in $(n,t)$. As a result, the right side of \eqref{eq:s_n1} is finite for any $u_0 \in L^2(\Omega_0)$. Since the bound on the partial sum is 
\beq\label{eq:ps_case1+2}
\norm{S_N(t)}_{L^2(\Omega_0)}^2  \leq C_j^2 \| u_0 \|_{L^2(\Omega_0)}^2, ~~~j=1,2,
\eeq
 the $L^2$ limit exists. For case 3, the cosine factor in the exponential of \eqref{eq:case3} 
is positive since 
$0 \leq \frac{\beta}{\alpha} < 1$. Since the eigenvalues $\lambda_n$ are increasing functions of $n$, the initial state $u_0$ has to be chosen so that
$$
\sum_{n} |u_{0,n}|^2  (1+ t^\alpha \lambda_n)^{2 \left( \frac{1-\beta}{\alpha} \right)} e^{2t \lambda_n^{\frac{1}{\alpha}}\cos \theta(\alpha,\beta)},
$$
is finite.  A sufficient condition for this is that $\{ n \in \N ~|~ u_{0,n} \neq 0 \}$  is finite, that is, we require $u_0 \in C_{F}(\Omega_0)$, the set of initial conditions $u_0 \in L^2(\Omega_0)$ with finitely-many nonzero Fourier coefficients $u_{0,n}$. Let $n^*$ denote the largest index for which $u_{0,n} \neq 0$. In this case, we obtain
\beq\label{eq:ps_case3}
\norm{S_N(t)}_{L^2(\Omega_0)}^2  \leq C_3^2 (1+ t^\alpha \lambda_{n^*})^{2 \left( \frac{1-\beta}{\alpha} \right)} e^{2t \lambda_{n^*}^{\frac{1}{\alpha}}\cos \theta(\alpha,\beta)}  \| u_0 \|_{L^2(\Omega_0)}^2. 
\eeq
The $L^2$ limit exists but the norm increases exponentially in time.
It follows from this, that for cases 1 and 2, with $u_0 \in L^2(\Omega_0)$, or case 3, with $u_0 \in C_F(\Omega_0)$, we have $u(t)$ in \eqref{eq4} exists for all $t > 0$. In addition, the analyticity properties of the Mittag-Leffler functions ensure that $u \in \cC(\overline{\R^+},L^2(\Omega_0))$.

\noindent
3. We next show that $u(t) \in D(H)$ provided $u_0$ satisfies the conditions in Theorem \ref{thm:existence1}.
For case 1, if $t > 0$, the first bound of Lemma \ref{lm1} 
shows that 
\begin{eqnarray}\label{eq:case1_DH}
\norm{H u(\cdot,t)}_{L^2(\Omega_0)}^2 & = & \sum_{n=1}^{\infty} \lambda_n^2 \abs{u_{0,n}}^2 \abs{E_{\alpha,1}(-i \lambda_n t^\alpha)}^2 \nonumber \\
& \leq & C^2 \sum_{n=1}^{\infty} \abs{u_{0,n}}^2 \frac{\lambda_n^2}{(1+\lambda_n t^\alpha)^2} \leq C^2 t^{-2 \alpha} \norm{u_0}_{L^2(\Omega_0)}^2. 
\end{eqnarray}
For case 2, we have $u_0 \in D(H)$, so that for any $t > 0$,
\begin{eqnarray}\label{eq:case2_DH}
\norm{H u(\cdot,t)}_{L^2(\Omega_0)}^2 & = & \sum_{n=1}^{\infty} \lambda_n^2 \abs{u_{0,n}}^2 \abs{E_{\alpha,1}(-i \lambda_n t^\alpha)}^2 \nonumber \\
& \leq & C_2^2 \norm{u_0}_{D(H)}^2. 
 \end{eqnarray}
For case 3, we have $u_0 \in C_F(\Omega_0)$, and since $C_F(\Omega_0) \subset D(H)$, we have for any $t > 0$,
\beq\label{eq:case3_DH}
\norm{H u(\cdot,t)}_{L^2(\Omega_0)}^2 \leq 
 C_3^2 (1+ t^\alpha \lambda_{n^*})^{2 \left( \frac{1-\beta}{\alpha} \right)} e^{2t \lambda_{n^*}^{\frac{1}{\alpha}}\cos \theta(\alpha,\beta)}  \| u_0 \|_{D(H)}^2.
 \eeq
As a consequence we have $Hu \in \cC(\R^+,L^2(\Omega_0))$, and hence $u \in \cC(\R^+, D(H))$.

\noindent
4. We next prove that $u(t)$ solves the TFSE. For all $\lambda>0$, we have
$$ 
\partial_t^\alpha E_{\alpha,1}((-i)^\beta \lambda t^\alpha)=(-i)^\beta \lambda E_{\alpha,1}((-i)^\beta \lambda t^\alpha),\ t >0.$$
Therefore, in light of \eqref{eq:case1_DH}, we obtain for all $t \in \R^+$, that
\begin{eqnarray*}
\partial_t^\alpha u(\cdot,t) &=& (-i)^\beta \sum_{n=1}^{\infty} \lambda_n u_{0,n} E_{\alpha,1}((-i)^\beta \lambda_n t^{\alpha}) \phi_n\\
& = & ( -i)^\beta H u(\cdot,t).
\end{eqnarray*}
Therefore, for $t \in \R^+$, the function $u(x, y, t)$ is a solution to \eqref{eq1} in $L^2(\Omega_0) \cap D(H)$ and $\partial_t^\alpha u \in \cC(\R^+,L^2(\Omega_0))$.

\noindent 
5. We establish the property that $u(t)$ converges to $u_0$ as $t \rightarrow 0$. 
For all $t \in \R^+$, we have
\begin{eqnarray}
\norm{u(\cdot,t)-u_0}_{L^2(\Omega_0)}^2 & = & \sum_{n=1}^{\infty} \abs{u_{0,n}}^2 \abs{E_{\alpha,1}(-i \lambda_n t^\alpha)-1}^2 \nonumber \\
& \leq &  \sum_{n=1}^{\infty} \abs{u_{0,n}}^2 \left( \abs{E_{\alpha,1}(-i \lambda_n t^\alpha)} + 1 \right)^2 \nonumber \\
& \leq & C(t) \norm{u_0}_{L^2(\Omega_0)}^2, \label{eq8} 
\end{eqnarray}
where $C(t) > 0$ is a constant in cases 1 and 2, and given by the expression on the right side of \eqref{eq:case3_DH} in case 3. Since $\lim_{t \downarrow 0} \left( E_{\alpha,1}(-i \lambda_n t^\alpha) - 1 \right)=0$, for all $n \in \N$, inequality \eqref{eq8} and the dominated convergence theorem yield that $\lim_{t \downarrow 0} \norm{u(\cdot,t)-u_0}_{L^2(\Omega_0)}=0$.
As a consequence, $u$ given by \eqref{eq4} is a solution to \eqref{eq1}--\eqref{eq3}.

\noindent \emph{Part 2: \textit{Uniqueness}}.


\noindent
6. In order to show that this  weak solution is unique, we assume that $u_0=0$ and prove that \eqref{eq1}--\eqref{eq3} admits only the trivial solution. To do so,
we take the scalar product  in $L^2(\Omega_0)$ of both sides of \eqref{eq1}, with $\phi_n$, for some $n \in \N$. Putting $u_n(t):=\langle u(\cdot,t) , \phi_n \rangle_{L^2(\Omega_0)}$, we get that
$$
 -i^\beta \partial_t^\alpha u_n(t) + \lambda_n u_n(t)=0. 
$$
Moreover, we have $\lim_{t \downarrow 0} u_n(t)=\langle \lim_{t \downarrow 0} u(\cdot,t), \phi_n \rangle_{L^2(\Omega_0)}=0$ from the expansion \eqref{eq4}.
Thus, $u_n(t)=0$ for a.e. $t \in \R^+$.  
This shows that $u(\cdot,t)=\sum_{n \geq 1} u_n(t) \phi_n=0$ in $L^2(\Omega_0)$ for all $t \in \R^+$.
\end{proof}


\section{Calculation of the edge current}\label{sec:current1}
\setcounter{equation}{0}

We consider the initial-value problem for the general TFSE with exponents 
$(\alpha,\beta) \in (0,1]$:
\beq\label{eq:ftse1}
(-i^\beta \partial_t^\alpha + H(\textbf{a}))u(x,y, t) = 0,
\eeq
with initial condition $u(x,y,t=0) = u_0(x,y)$. 
The solution is given in terms of the Mittag-Leffler functions:
\beq\label{eq:solution1}
u(\cdot,t) = (E_{\alpha,1}((-i)^\beta t^\alpha H(\textbf{a})) u_0) (\cdot), 
\eeq
where $(\cdot)$ stands for the spatial coordinates $(x,y) \in \Omega := \R_x^+ \times \R_y$.
To compute the edge current $J_y[u_0]$, we need 
\beq\label{deriv_soln1}
\partial_t u(\cdot,t) = (-i)^\beta t^{\alpha - 1} (E_{\alpha,\alpha}((-i)^\beta t^\alpha H(\textbf{a}) ) H(\textbf{a}) u_0)(\cdot),
\eeq 
where we used the identity \eqref{eq:deriv_ml1} for the derivative of the Mittag-Leffler function.  
This provides a compact expression for the edge current: 
\bea\label{eq:Gcurrent1}
J_y[u_0](t) &:= &\partial_t \langle y u(\cdot, t) , u(\cdot, t) \rangle_{L^2(\Omega)} \nonumber \\
 &=& 2 t^{\alpha -1} \Re \{ (-i)^\beta \langle y E_{\alpha,\alpha}((-i)^\beta t^\alpha H(\textbf{a})) H(\textbf{a}) u_0, 
    E_{\alpha,1}((-i)^\beta t^\alpha H(\textbf{a})) u_0  \rangle_{L^2(\Omega)} \} . 
 \nonumber \\
 & &
\eea
\subsection{Choice of the initial state $u_0$}
We select an initial state that is a truncation in the $k$-variable of the first eigenfunction $\phi_1(x,k)$ of the fiber operator $h_b(k)$ with eigenvalue $\lambda_1(k)$. We choose a compactly supported function $\chi(k)$ with support in a closed interval $I \subset \R$ so that $\lambda_1(I) \subset (b, 3b)$.  
We select $u_0(x,y)$ by taking 
\beq\label{eq:initial1}
\widehat{u_0}(x,k) = \frac{1}{\sqrt{2 \pi}} \int_\R e^{-iky} u_0(x,y) ~dy := \chi(k) \phi_1(x,k).
\eeq
We recall that we choose $\phi_1(x,k)$ to be real. 


\subsection{Formula for the edge current} 

To facilitate the calculation of $J_y[u_0]$, we define two functions
\bea\label{eq:funct1}
u_\alpha  &:=  & E_{\alpha,1}((-i)^\beta t^\alpha H(\textbf{a})) u_0 \\
v_\alpha  &:= &  E_{\alpha,\alpha}((-i)^\beta t^\alpha H(\textbf{a})) u_0 .
\eea
Using these two functions, we may express the current in \eqref{eq:Gcurrent1} as
\beq\label{eq:Acurrent2}
J_y [u_0](t) =  2 t^{\alpha - 1}  \Re \{  (-i)^\beta \langle H(\textbf{a}) v_\alpha , y u_\alpha \rangle \} .
\eeq

Let  us reduce the matrix element
\beq\label{eq:k_term1.12}
K_{u_0}(t) := \langle H(\textbf{a}) v_\alpha , y u_\alpha \rangle. 
\eeq
Using the choice of $u_0$ in \eqref{eq:initial1}, we obtain for the partial Fourier transforms with respect to $y$:
\beq
\widehat{u_\alpha}(x,k,t) =    E_{\alpha, 1}((-i)^\beta t^\alpha \lambda_1(k)) \chi(k) \phi_1(x,k),
\eeq
and
\beq
\widehat{v_\alpha}(x,k,t) =    E_{\alpha, \alpha}((-i)^\beta t^\alpha \lambda_1(k)) \chi(k) \phi_1(x,k) .
\eeq
Substituting these into the expression \eqref{eq:k_term1.12} for $K_{u_0}(t)$ yields
\beq\label{eq:k2.1}
K_{u_0}(t) =  \int_\R ~\langle h_b(k) \widehat{v_\alpha}(\cdot,k,t), \widehat{y u_\alpha}(\cdot,k,t) \rangle_{L^2(\R^+)}  ~dk .
\eeq

We note that 
\beq\label{eq:derivML1.1}
\widehat{y u_\alpha}(x,k,t) = i \partial_k ( E_{\alpha, 1}((-i)^\beta t^\alpha \lambda_1(k)) \chi(k) \phi_1(x,k) ).
\eeq
We obtain from \eqref{eq:k2.1}-\eqref{eq:derivML1.1}:
\bea\label{eq:k3.1}
K_{u_0}(t) & = &  \int_\R  \lambda_1(k) \chi(k)  E_{\alpha, \alpha}((-i)^\beta t^\alpha \lambda_1(k))  
     \langle   \phi_1(\cdot,k),  \nonumber \\
    & &   i \partial_k( \chi(k) \phi_1(\cdot,k) E_{\alpha, 1}((-i)^\beta t^\alpha \lambda_1(k)) ) \rangle_{L^2(\R^+)}  ~dk.
  \eea


The inner product in \eqref{eq:k3.1} may be expanded to obtain
\bea
\lefteqn{\langle \phi_1(\cdot,k), i\partial_k( E_{\alpha, 1}((-i)^\beta t^\alpha \lambda_1(k)) \chi(k) \phi_1(\cdot,k) ) \rangle_{L^2(\R^+)} } \nonumber \\
 &=&  - i \partial_k ( \overline{E_{\alpha, 1}((-i)^\beta t^\alpha \lambda_1(k))} \chi(k)) ,
\eea
since  $\langle \phi_1(\cdot,k), \phi_1(\cdot,k) ) \rangle_{L^2(\R^+)} =1$,
and $\langle \phi_1(\cdot,k), \partial_k \phi_1(\cdot,k) ) \rangle_{L^2(\R^+)} =0$, since $\phi_1(x,k)$ may be chosen to be real.
Consequently,  the expression for $K_{u_0}(t)$ is
\beq\label{eq:k5.1}
K_{u_0}(t)  =  -i \int_\R  \lambda_1(k)  E_{\alpha, \alpha}((-i)^\beta t^\alpha \lambda_1(k)) \chi(k) \partial_k (\overline{E_{\alpha, 1}((-i)^\beta t^\alpha \lambda_1(k)) } \chi(k) ) ~ dk .
\eeq

We write $E_{\alpha, j}(k,t)  := E_{\alpha,j}((-i)^\beta t^\alpha \lambda_1(k))$, 
for $j = 1, \alpha$, for short. 
We decompose $K_{u_0}(t) := -i[ K_{u_0}^{(1)}(t) +  K_{u_0}^{(2)}(t)]$, where
\beq\label{eq:k_1.1}
 K_{u_0}^{(1)}(t) := \int_\R  \lambda_1(k) \chi (k) \chi^\prime(k) E_{\alpha,\alpha}(k,t) \overline{E_{\alpha,1}(k,t)}  ~ dk ,
 \eeq
 and
\beq\label{eq:k_2.1}
 K_{u_0}^{(2)}(t) :=   \int_\R  \lambda_1(k) \chi^2 (k)  E_{\alpha,\alpha}(k,t) 
 \partial_k \overline{E_{\alpha,1}(k,t)}  ~ dk .
 \eeq
 We may reduce the form of $ K_{u_0}^{(2)}(t)$ using the fact that 
 \beq\label{eq:Ederiv1.1}
 \partial_k {E_{\alpha,1}(k,t)} = \frac{1}{\alpha} (-i)^{\beta} t^\alpha \lambda_1^\prime(k) E_{\alpha, \alpha}(k,t).
 \eeq
 From this, we may write 
 \beq\label{eq:k_22.2}
  K_{u_0}^{(2)}(t) := \frac{i^\beta t^\alpha}{\alpha} \int_\R  \lambda_1(k) \lambda_1^\prime(k) \chi^2 (k) | E_{\alpha,\alpha}(k,t) |^2  ~dk .
 \eeq
 
 The final formula for the current is:
 \bea\label{eq:courantF}
 J_y [u_0](t) & = & 2 t^{\alpha - 1}  \Re \left\{  (-i)^{1+\beta} 
       \int_\R \left[   \lambda_1(k) \chi (k) \chi^\prime(k) E_{\alpha,\alpha}(k,t) \overline{E_{\alpha,1}(k,t)} \right. \right.   \nonumber \\
  &   &  \left. \left.  + i^\beta \frac{t^\alpha}{\alpha}  \lambda_1(k) \lambda_1^\prime(k) \chi^2 (k) | E_{\alpha,\alpha}(k,t) |^2
  \right] ~ dk  \right\}  
  \nonumber \\
   & = & 2 t^{\alpha - 1}  
       \int_\R  \lambda_1(k) \chi (k) \chi^\prime(k) \Re \left\{  (-i)^{1+\beta}  E_{\alpha,\alpha}(k,t) \overline{E_{\alpha,1}(k,t)} \right\} ~dk  . 
        \nonumber \\
        & &
\eea

\medskip

\begin{remark}\label{rmk:schr_case1}
The formula \eqref{eq:courantF} reduces to the classical result when $\alpha = \beta = 1$:
\beq\label{eq:courantCl1}
J_y [u_0](t) =  - 2  \left\{  \int_\R    \lambda_1(k) \chi (k) \chi^\prime(k) 
|E_{1,1}(k,t)|^2  ~dk  \right\} . 
\eeq
We recall that $E_{1,1}(z) = e^z$ so that $E_{1,1}(k,t)= e^{-i \lambda_1(k) t}$, and hence
\bea\label{eq:courantCl2}
J_y [u_0](t) &= & - 2  \left\{  \int_\R    \lambda_1(k) \chi (k) \chi^\prime(k) 
 ~dk  \right\}  \nonumber \\
  &=& \int_\R  \lambda_1^\prime (k) \chi^2 (k)  ~dk   . 
\eea
This result agrees with the calculation of the edge current for the \Schr equation using the unitary group  since  $u(t) = e^{-i H({\bf{a}})t} u_0$ so that one may use \eqref{eq:current2}, with initial condition \eqref{eq:initial1}. Explicit positive lower bounds on $\lambda_1^\prime(k)$ are given in \cite[Theorem 2.1]{hs1}. 
\end{remark}

\medskip

\begin{remark}\label{rmk:schr_case2}
It is interesting to compute the edge current for $\alpha =1$ and $\beta \in (0,1)$. In this case, the solution to the initial-value problem is similar to the \Schr case: $u(t) = e^{(-i)^\beta H({\bf{a}})t} u_0$. Using \eqref{eq:courantF}, with the fact that $E_{1,1}(k,t)= e^{(-i)^\beta \lambda_1(k) t}$, we find that the edge current has the form:
\beq\label{eq:courant_beta1}
{J_y} [u_0](t) = 2 \cos \left( \frac{\pi}{2} ( 1 + \beta) \right)
       \int_\R  \lambda_1 (k) \chi^\prime (k) \chi (k) e^{2t \lambda_1(k)\cos \left( \frac{\pi \beta}{2} \right)}  ~dk.
       \eeq
We note that when $\beta =1$, we obtain \eqref{eq:courantCl2}.
\end{remark}


 \section{Asymptotic behavior of the edge current}\label{sec:asympt1}
\setcounter{equation}{0}

In this section, we use formula \eqref{eq:courantF} of section \ref{sec:current1} in order to calculate the asymptotic behavior in time of the edge currents.
We first summarize the asymptotic expansions of the Mittag-Leffler functions.

 \subsection{Asymptotic expansions of Mittag-Leffler functions}
 
 In order to analyze the long-time behavior of $J_y[u_0](t)$, we need the large $t$ behavior of the functions $E_{\alpha,\alpha}(k,t)$ and $E_{\alpha,1}(k,t)$. 
We use formulas \eqref{eq:asymp1}-\eqref{eq:asymp2} of Appendix \ref{subsec:MLasymptotics1} to write the expansions of $E_{\alpha,\sigma}(k,t)$ to fourth order in $\kappa$. For this, we recall the notation:
 $$
 E_{\alpha, \sigma}(k,t)  := E_{\alpha,\sigma}((-i)^\beta t^\alpha \lambda_1(k)) ,
 $$ 
for $\sigma = 1, \alpha$. The choice of the expansion depends on the relationship between $\alpha$ and $\beta$. To apply the asymptotics of the Mittag-Leffler functions, we note that $z = (-i)^\beta \kappa$, with $\kappa := t^\alpha \lambda_1(k) > 0$, so that $| \arg z| = \frac{\pi \beta}{2} < \pi$, for $\beta \in [0,1]$. 
The Mittag-Leffler functions have different asymptotic behaviors in $z$ depending on the sector in which $|z| \rightarrow \infty$. These sectors are determined by a real parameter $\mu$ satisfying \eqref{eq:mu1}. We have two cases:

\medskip

\begin{enumerate}
\item[\textbf{Case 1}:] $\beta \leq \alpha$. $|\arg z| < \mu$ for all $\mu$ satisfying \eqref{eq:mu1}. The asymptotics \eqref{eq:asymp1} hold.

\medskip

\item[\textbf{Case 2:}] $\alpha < \beta$. All $\mu \in  \left(\frac{\pi \alpha}{2}, \min (\pi \alpha, \frac{\pi \beta}{2} )  \right)$ satisfy \eqref{eq:mu1}, and we have $\mu < |\arg z| < \pi$. The asymptotics \eqref{eq:asymp2} hold.

\end{enumerate}

%
%
%
%
%
%
From this analysis, the fact that $z = (-i)^\beta \kappa$, with $\kappa := t^\alpha \lambda_1(k) > 0$, and the asymptotics \eqref{eq:asymp1}-\eqref{eq:asymp2}, we obtain:

\medskip

\noindent
\textbf{Case 1.} ${\beta \leq \alpha}$. 
\begin{align}
& E_{\alpha,1}(z) =  \frac{1}{\alpha} e^{(-i)^{\frac{\beta}{\alpha}}\kappa^{\frac{1}{\alpha}}} - \frac{i^{\beta}}{\kappa \Gamma (1 - \alpha)} - 
\frac{i^{2\beta}}{\kappa^2 \Gamma (1 - 2 \alpha)} - \frac{i^{3\beta}}{\kappa^3 \Gamma (1 - 3 \alpha)}  + \mathcal{O}(\kappa^{-4})  \label{eq:a1.1}\\
 & \nonumber \\
 & E_{\alpha,\alpha}(z) = 
 \frac{1}{\alpha} (-i)^{\frac{\beta ( 1- \alpha)}{\alpha}} \kappa^\frac{1-\alpha}{\alpha} e^{(-i)^{\frac{\beta}{\alpha}}\kappa^{\frac{1}{\alpha}}} - 
 \frac{i^{2\beta}}{\kappa^2 \Gamma ( - \alpha)} -\frac{i^{3\beta}}{\kappa^3 \Gamma ( - 2 \alpha)}  + \mathcal{O}(\kappa^{-4}) 
 \label{eq:a1.2} 
  \end{align}
  
  \medskip

\noindent
\textbf{Case 2.} ${\beta > \alpha}$. 
\begin{align}
&\label{eq:2.1}
E_{\alpha,1}(z) =   - \frac{i^{\beta}}{\kappa \Gamma (1 - \alpha)} - \frac{i^{2\beta}}{\kappa^2 \Gamma (1 - 2\alpha)} -\frac{i^{3\beta}}{\kappa^3 \Gamma (1 - 3\alpha)}  + \mathcal{O}(\kappa^{-4})  \\
& \nonumber \\
& \label{eq:2.2}
E_{\alpha,\alpha}(z) = - \frac{ i^{2\beta}}{\kappa^2 \Gamma(- \alpha)} - \frac{ i^{3\beta}}{\kappa^3 \Gamma(-2 \alpha)}  
  + \mathcal{O}(\kappa^{-4})
\end{align} 
We note that $\Gamma(-n)^{-1} = 0$, for $n \in \N \cup \{ 0 \}$.
 
We recall that the integrand of the edge current contains the factor:
\beq\label{eq:real1}
 \Re \left\{  (-i)^{1+\beta} 
        E_{\alpha,\alpha}(k,t) \overline{E_{\alpha,1}(k,t)}  \right\},
\eeq
whose asymptotic expansion we now compute for each case.  We note that as long as $\alpha, \sigma \geq  0$, the Mittag-Leffler function satisfies
 $$
 \overline{E_{\alpha,\sigma} (z)} = E_{\alpha,\sigma} (\overline{z}) .
 $$

\subsection{Asymptotic expansion for the current in Case 1: $0 < \beta \leq \alpha < 1$.}

We first compute the leading term of  \eqref{eq:real1} using the asymptotics \eqref{eq:a1.1}-\eqref{eq:a1.2}. 
The result is
\beq\label{eq:leading1.11}
\frac{1}{\alpha^2} \cos \left[ \left( \frac{\pi \beta}{2 \alpha}  \right)(1-\alpha) + \frac{\pi}{2}(1 + \beta) \right]   \lambda_1(k)^\frac{1-\alpha}{\alpha} t^{1 - \alpha} e^{2 t \lambda_1(k)^\frac{1}{\alpha} \cos(\frac{\pi \beta}{2 \alpha} ) } .
\eeq
This term is positive since $0 < \frac{\beta}{\alpha} \leq 1$. The leading term grows exponentially in time. 
The other terms in the expansion are of lower order. To simplify the notation, we introduce two angles:
\beq\label{eq:angles1}
\theta := \frac{\pi \beta}{2 \alpha} , ~~~~ \gamma (k):=  \lambda_1(k)^\frac{1}{\alpha} \sin \theta . 
\eeq
We express the next lower-order term in \eqref{eq:real1} as follows:
\beq\label{eq:leading1.12}
- t^{1 - 2 \alpha}  e^{ t \lambda_1(k)^\frac{1}{\alpha} \cos \theta } 
 \left[ \frac{1}{\alpha \Gamma(1-\alpha)} \cos \left(t \gamma (k) + \theta +\frac{\pi}{2} (1+\beta) \right)    \lambda_1(k)^\frac{1-2\alpha}{\alpha} \right]  .
\eeq
The remainder is seen to be $\mathcal{O}(t^{-2 \alpha}).$

As a consequence of \eqref{eq:leading1.11}-\eqref{eq:leading1.12}, the edge current \eqref{eq:courantF} may be expressed as 
 \bea\label{eq:edge_case1}
  J_y [u_0](t) & = &  2 t^{\alpha - 1}  \Re \left\{  (-i)^{1+\beta} 
       \int_\R \left[   \lambda_1(k) \chi (k) \chi^\prime(k) E_{\alpha,\alpha}(k,t) \overline{E_{\alpha,1}(k,t)} \right] ~dk  \right\}  \nonumber \\
        & = & \frac{2}{\alpha^2} \cos \left( \theta (1 - \alpha) + \frac{\pi}{2}(1 + \beta) \right)   \int_\R \lambda_1^\frac{1}{\alpha}(k) \chi(k) \chi^\prime(k) e^{2 t \lambda_1^\frac{1}{\alpha}(k) \cos \theta } ~dk \nonumber           \\
         & & -  \frac{2 t^{-\alpha} }{\alpha \Gamma(1-\alpha)} \int_\R  \cos \left( t\gamma(k)  + \theta +\frac{\pi}{2} (1+\beta) \right)      
         \lambda_1^\frac{1-\alpha}{\alpha}(k) \chi(k) \chi^\prime(k) 
         e^{ t \lambda_1^\frac{1}{\alpha}(k) \cos \theta}   ~dk \nonumber           \\
          & & + \mathcal{O}(t^{-1-\alpha} ).
 \eea       
 
The main characteristic of the edge current in Case 1 with $0 < \beta < \alpha < 1$ is the exponential growth of the current with time. However, for the case $\alpha = \beta$, the  factor in the exponential $\cos \theta = 0$ and the edge current is asymptotically constant in time, see section \ref{subsec:naber1}.


\subsection{Asymptotic expansion for the current in Case 2: $0 < \alpha < \beta \leq 1$.}

The edge current in Case 2 decays with time. 
Using the asymptotics for Case2, \eqref{eq:2.1}-\eqref{eq:2.2}, we find:
 \bea\label{eq:edge_case2}
 J_y [u_0](t) & = & \frac{2}{t^{1+ 3\alpha}} \left( \int_\R \frac{1}{\lambda_1^3(k)} \chi(k) \chi^\prime(k) ~dk \right)  \nonumber \\
  & & \times \cos \left( \frac{(\beta + 1) \pi}{2} \right) \left[ \frac{1}{\Gamma(1-2\alpha) \Gamma(-\alpha)} - \frac{1 }{ \Gamma(1-\alpha) \Gamma(-2 \alpha)} \right]  \nonumber \\
  & & + \mathcal{O}(t^{-1 - 4 \alpha}) .
  \eea
In the regime of Case 2, $0 < \alpha < \beta \leq 1$, the edge current decays like 
$t^{- 1- 3\alpha}$.


\section{Asymptotics of the edge current in special cases}\label{eq:special1}
\setcounter{equation}{0}

In this section, we discuss the results of section \ref{sec:asympt1} for two well known models.

\subsection{Asymptotics of the current for the Naber model}\label{subsec:naber1}

The original TFSE model proposed by Naber \cite{naber} consists of taking $0 < \alpha = \beta < 1$. This model is an example of Case 1. We note that for $\alpha = \beta > 0$, the term in the exponent $(-i)^{\frac{\beta}{\alpha}} = -i$ and the exponential factors in \eqref{eq:a1.1} and \eqref{eq:a1.2} have modulus one. Consequently, the edge current \eqref{eq:edge_case1} in this case has the expansion: 
 \bea\label{eq:edge_case1N}
  J_y [u_0](t) & = &   2 t^{\alpha - 1}  \Re \left\{  (-i)^{1+\alpha} 
       \int_\R \left[   \lambda_1(k) \chi (k) \chi^\prime(k) E_{\alpha,\alpha}(k,t) \overline{E_{\alpha,1}(k,t)} \right] ~dk  \right\}  \nonumber \\
        & = & \frac{1}{\alpha^2} \int_\R (\lambda_1(k)^\frac{1}{\alpha})^\prime \chi^2(k)  ~dk \nonumber           \\
      & & + \frac{2t^{-\alpha}}{\alpha \Gamma(1-\alpha)} \int_\R  \lambda_1^{\frac{1- \alpha}{\alpha} }(k) \chi(k) \chi'(k) \cos \left( \frac{ \pi \alpha}{2} + t \lambda_1^{\frac{1}{\alpha}}(k) \right) ~dk  \nonumber \\
       & & + \mathcal{O}(t^{-1 - \alpha}) . 
      \eea
Thus, the edge current in the Naber model is a asymptotically constant in time.        
Taking $\alpha = \beta =1$, the leading term of \eqref{eq:edge_case1N} is the same as \eqref{eq:courantCl2} for the \Schr model, and the next term in the expansion vanishes. 
         

\subsection{Asympotics of the current for the AYH model}\label{subsec:achar1}

The TFSE model proposed by Narahari Achar, et.\ al., \cite{ayh} consists of taking $0 < \alpha < 1$ and $\beta = 1.$ This is an example of Case 2. 
Evaluating the asymptotic formula in \eqref{eq:edge_case2}, we obtain
\bea\label{eq:edge_achar1}
 J_y [u_0](t) & = & - \frac{2}{t^{1+ 3\alpha}} \left( \int_\R \frac{1}{\lambda_1^3(k)} \chi(k) \chi^\prime(k) ~dk \right) \left[ \frac{1}{\Gamma(1-2\alpha) \Gamma(-\alpha)} - \frac{1}{ \Gamma(1-\alpha) \Gamma(-2 \alpha)} \right]  \nonumber \\
  & & + \mathcal{O}(t^{-1 -4 \alpha}) .
  \eea
The edge current decays to zero as $t \rightarrow \infty$.  We note that, unlike the Naber model, when $\alpha \rightarrow 1^-$, the leading term in \eqref{eq:edge_achar1} vanishes.

\section{Mean square displacement in the time-fractional Hall model}\label{sec:msd1}

\setcounter{equation}{0}


In a related paper \cite{hs_msd1}, we studied the effect of the fractional time derivative on the free propagation properties of wave packets in $\R^d$, $d \geq 1$, through the evolution of mean square displacement. 
As another indicator of the anomalous transport induced by a fractional time derivative in the quantum Hall model, we compute the mean square displacement (MSD) of a wave packet in the $y$ direction for the TFSE: 
\beq\label{eq:tfse1}
i^\beta \partial_t^\alpha u = H({\bf{a}}) u, ~~~ u(t=0) = u_0,
\eeq
with an initial condition $u_0$ as defined in \eqref{eq:initial1}. 

We recall that the solution to \eqref{eq:tfse1} $u(x,y,t)$, with $(x,y) \in \Omega := \R_x^+ \times \R_y$, is given by
\beq\label{eq:tfse_soln1}
u(x,y,t) = E_{\alpha,1}( (-i)^\beta t^\alpha H({\bf{a}})) u_0(x,y),
\eeq
so that after a partial Fourier transform with respect to the $y$ variable, we have
\beq\label{eq:tfse_ft1}
\widehat{u}(x,k,t)  = E_{\alpha,1}( (-i)^\beta t^\alpha \lambda_1(k)) \chi(k) \phi_1(x,k) . 
\eeq

We define the MSD for the solution $u(x,y,t)$ in \eqref{eq:tfse_soln1} by
\beq\label{eq:msd1}
D_2(u_0,t) := \langle u(t), y^2 u(t) \rangle_{L^2(\Omega)} . 
\eeq
Similar to calculations in section \ref{sec:current1}, by means of \eqref{eq:derivML1.1}, we easily find
\beq\label{eq:msd2}
D_2(u_0,t) = \int_\R  |\partial_k\widehat{u}(x,k,t) |^2 ~dx ~dk , 
\eeq
where, using \eqref{eq:tfse_ft1} and \eqref{eq:deriv_ml1} for the derivative of the Mittag-Leffer function, we obtain, 
\beq\label{eq:msd3}
\partial_k\widehat{u}(x,k,t) = I + II + III .
\eeq
Using the notation $E_{\alpha,j }(k,t) := E_{\alpha,j}( (-i)^\beta t^\alpha \lambda_1(k))$, for $j = 1, \alpha$, as above, we have for the three terms in \eqref{eq:msd3}:
\bea\label{eq:tfse_ft2}
I & = & E_{\alpha,\alpha}(k,t ) (-i)^\beta t^\alpha \lambda^\prime_1(k) \chi(k) \phi_1(x,k) , \nonumber \\
II & = &  E_{\alpha,1}( k,t) \chi^\prime(k) \phi_1(x,k) , \nonumber \\
III & =&  E_{\alpha,1}(k,t)  \chi(k) \partial_k \phi_1(x,k) . 
\eea
Using the properties of the eigenfunction $\phi_1(x,k)$, and its orthogonality to $\partial_k \phi_1 (x,k)$, we find that the MSD is the sum of four terms
$$
D_2(u_0,t) = A + B + C +F,
$$
where the square terms are 
\bea\label{eq:msd5}
A &=& \int_\R | E_{\alpha,\alpha}(k,t )|^2  t^{2\alpha} ( \lambda^\prime_1(k))^2 \chi^2(k) ~dk , \nonumber \\
 B &=& \int_\R | E_{\alpha, 1}(k,t )|^2  (\chi^\prime(k))^2 ~dk ,  \nonumber \\
 C & = & \int_\R | E_{\alpha, 1}(k,t )|^2   \chi^2(k) \left( \int_{\R^+} | \partial_k \phi_1(x,k)|^2 dx \right) ~dk ,
\eea
and the one nonvanishing cross term is
\beq\label{eq:msd6}
F= 2 t^\alpha \int_\R \Re \left\{  (-i)^\beta  E_{\alpha,\alpha}(k,t ) \overline{ E_{\alpha,1}(k,t )}  \right\}  \lambda^\prime_1(k) \chi^\prime (k) \chi(k) ~dk .
\eeq


\noindent \emph{Case 1.} For $0 < \beta < \alpha < 1$, the asymptotics \eqref{eq:a1.1}-\eqref{eq:a1.2} indicate that the MSD increases exponentially in time. However, for the case $0 < \alpha = \beta < 1$, the Naber model, the asymptotic behavior of the MSD is a result of the asymptotics given in \eqref{eq:a1.1}-\eqref{eq:a1.2}, for $\alpha = \beta$: 
\beq\label{eq:msd1equal}
D_2(u_0,t) = \frac{t^2}{\alpha^2} \int_\R \lambda_1(k)^{2\left(  \frac{1-\alpha}{\alpha}\right) } \lambda_1^\prime(k)^2 \chi_1^2(k) ~ dk + \mathcal{O}(1).
\eeq
Hence,  the MSD $D_2(u_0,t)$ is ballistic in the $y$-direction for the model of Naber \eqref{eq:naber0} (with $-\Delta$ replaced by $H({{\bf a}})$, and $q=0$). 

\medskip

\noindent \emph{Case 2: $0< \alpha < \beta \leq 1$}. The expansions \eqref{eq:2.1}-\eqref{eq:2.2} lead to the following expression for the asymptotics of the MSD:
\bea\label{eq:msd4}
D_2(u_0,t) & = &  \frac{1}{t^{2 \alpha}} \left\{  \frac{1}{\Gamma(-\alpha)^2} \int_\R \frac{ \lambda^\prime (k)^2}{\lambda_1^4(k)} \chi^2(k) ~dk \right. \nonumber \\
 & &  \left. + 
\frac{1}{\Gamma(1-\alpha)^2} \int_\R \frac{ ( \chi^\prime (k)^2 + \chi^2(k) \Phi(k) ) }{\lambda_1^2(k)} \chi^2(k) ~dk \right\} + \mathcal{O}(t^{-3\alpha}) ,
   \nonumber \\
  \eea
where 
$$
\Phi(k) := \int_{\R^+} | \partial_k \phi_1(x,k)|^2 dx   .
$$
In this case, the MSD in the $y$-direction decays like $t^{- 2 \alpha}$.

In summary, the MSD in the $y$-direction behaves like $t^2$ for the Naber model and like $t^{-2 \alpha}$ for the AYH model. For fixed $\alpha$, the MSD, like the edge current, exhibits a transport transition between $0< \beta < \alpha$ and $\alpha \leq \beta \leq 1$. 



\begin{appendices}

\section{Mittag Leffler functions}\label{app1:ml1}
\setcounter{equation}{0}

\subsection{Definitions}

The Mittag-Leffler functions are generalizations of the exponential function. They are indexed by two parameters $(\alpha, \sigma)$: 
\beq\label{eq:defn_ml}
E_{\alpha,\sigma}(z) := \sum_{n=0}^\infty \frac{z^n}{\Gamma (\alpha n + \sigma)} .
\eeq
These are entire functions of $z \in \C$ for $\alpha, \sigma > 0$. Note that $E_{1,1}(z) = e^{z}$. The function $E_{\alpha,1}(-\lambda t^\alpha)$ satisfies the fractional differential equation:
$$
(\partial_t^\alpha + \lambda) E_{\alpha,1}(-\lambda t^\alpha) = 0.
$$

We need the derivative of the Mittag-Leffler function $E_{\alpha,1}(z)$ with respect to $z$. 
\beq\label{eq:deriv_ml1}
\frac{d E_{\alpha,1}}{dz}(z) = \sum_{m=0}^\infty \frac{(m+1) z^m}{\Gamma(\alpha(m+1) +1)} = \frac{1}{\alpha} \sum_{m=0}^\infty \frac{z^m}{\Gamma(\alpha m+ \alpha)} =  \frac{1}{\alpha} E_{\alpha, \alpha}(z),
\eeq
using the property that $\Gamma ( z +1) = z \Gamma(z).$  
\subsection{Bounds on Mittag-Leffler functions}\label{subsec:MLbounds1}

The following technical result, needed by the proof of Theorem \ref{thm:existence1}, 
is a byproduct of \cite[(1.147),  (1.148)]{P}. The asymptotic expansions are presented in \eqref{eq:asymp1} and \eqref{eq:asymp2}. We consider only the case $\alpha \in (0, 1]$.

\begin{lemma}\cite[Theorems 1.5 and 1.6]{P}
\label{lm1}
Let $\sigma \in \R$, $\alpha < 2$, and $\mu \in \left(\frac{\pi \alpha}{2}, \min \{\pi \alpha, {\pi} \} \right)$.
\begin{enumerate}
\item There exists a finite constant $C=C(\alpha,\sigma)>0$  so that 
 $$
|E_{\alpha,\sigma}(z)| \leq  C \left( (1 + \abs{z})^{\frac{1-\sigma}{\alpha}}e^{\Re (z^{\frac{1}{\alpha}})} +\frac{1}{1 + \abs{z}} \right), 
$$
for all $|\arg z| \leq \mu$ and $|z| \geq 0$.

\medskip

\item There exists a finite constant $C=C(\alpha,\sigma)>0$  so that 
 $$
|E_{\alpha,\sigma}(z)| \leq  \frac{C}{1+|z|} ,
$$
for all $\mu < |\arg z| \leq \pi$ and $|z| \geq 0$.
\end{enumerate}

\end{lemma}

\subsection{Asymptotic expansions of the Mittag-Leffler functions}\label{subsec:MLasymptotics1}

The asymptotics of the Mittag-Leffler functions are presented in section 1.2.7 of Podnubny \cite{P}, Theorems 1.3 and 1.4. 
The asymptotics as $|z| \rightarrow \infty$ depend on the location of $z \in \C$.  For any $\alpha \in (0,1)$ (actually, $\alpha \in (0,2)$ is allowed),  and $\sigma \in \C$, choose any $\mu >0$ satisfying
 \beq\label{eq:mu1}
 \frac{\pi \alpha}{2} < \mu < \min \{ \pi , \pi \alpha \} = \pi \alpha. 
 \eeq
 Theorem 1.3 of section 1.2.7 of \cite{P} states that if $|\arg z| \leq \mu$, then, for any $p \geq 1$, we have:
 \beq\label{eq:asymp1}
 E_{\alpha,\sigma}(z) = \frac{1}{\alpha} z^{\frac{1-\sigma}{\alpha}}e^{z^{\frac{1}{\alpha}}} - \sum_{k=1}^p \frac{z^{-k}}{\Gamma (\sigma - \alpha k)} + \mathcal{O}(|z|^{-p-1}).
 \eeq
 Theorem 1.4 of section 1.2.7 of \cite{P} states that if $\mu \leq  |\arg z| \leq \pi$, then, for any $p \geq 1$, we have:
 \beq\label{eq:asymp2}
 E_{\alpha,\sigma}(z) = - \sum_{k=1}^p \frac{z^{-k}}{\Gamma (\sigma - \alpha k)} + \mathcal{O}(|z|^{-p-1}).
 \eeq
We note that $\Gamma(-n)^{-1} = 0$, for $n \in \N \cup \{ 0 \}$. 

\end{appendices}



\end{document}